\documentclass[pdflatex,sn-mathphys-num]{sn-jnl}

\usepackage{graphicx}%
\usepackage{multirow}%
\usepackage{amsmath,amssymb,amsfonts}%
\usepackage{amsthm}%
\usepackage{mathrsfs}%
\usepackage[title]{appendix}%
\usepackage{xcolor}%
\usepackage{textcomp}%
\usepackage{manyfoot}%
\usepackage{booktabs}%
\usepackage{algorithm}%
\usepackage{algorithmicx}%
\usepackage{algpseudocode}%
\usepackage{listings}%
\usepackage{enumerate}

\begin{document}

\title[Article Title]{On the concept of simultaneity in relativity} 


\author*[1]{\fnm{Justo Pastor} \sur{Lambare}}\email{jupalam@gmail.com}

\affil*[1]{\orgdiv{Facultad de Ciencias Exactas y Naturales}, \orgname{Universidad Nacional de Asunci\'{o}n}, \orgaddress{\street{Ruta Mcal. J. F. Estigarribia, Km 11 Campus de la UNA}, \city{San Lorenzo}, \country{Paraguay}}} 
\abstract{
In this comment, we demonstrate that the claim by Spavieri et al., asserting that Wang et al.'s interferometric experiment disproves the special theory of relativity by revealing that simultaneity must be an absolute concept independent of the observer's state of motion, 
is based on circular reasoning and therefore constitutes a logical fallacy.
}


\maketitle
%
\section{Introduction}\label{sec:intro}
%
In Ref. \cite{pSpa25}, the authors present several arguments against the validity of special relativity.
They follow an out-of-the-mainstream tradition started in the 1930s by Albert Eagle \cite{pEag38}, pretending to prove that absolute simultaneity is a natural concept imposed on us by rational thinking and experimental facts whose rejection leads to inconsistencies.
A quick literature search reveals that this anti-relativistic tradition remains alive in the 21st century \cite{pEng15,pSzo20,pLee20,pD&W20,pCho18,pKip21b,pSpa21,pGif25}.

One of the leading figures of this tradition was the Italian physicist Franco Selleri, who proposed an argument known among his followers as \textit{Selleri's paradox}.

The argument proposed by Selleri  \cite{pSel96,pSel97} never received serious consideration by mainstream physicists, which is probably unfortunate because the clarification of foundational issues has didactic value and restrains the spreading of incorrect scientific ideas among non-experts and the public in general.\footnote{A didactic refutation of Selleri's argument can be found in Ref.\cite{pLam24a}.}

Although all the arguments exposed by Spavieri et al. in Ref.\cite{pSpa25} can be consistently rebutted according to the orthodox scientific understanding, we shall concentrate only on clarifying their concrete claim stating that Wang et al.'s interferometric experiment (Wang experiment for short) cannot be consistently explained by special relativity.

The argument is sufficiently compelling as to be considered a paradox.
We congratulate Spavieri et al. for conceiving and promoting the paradox.
However, there is a difference between a paradox, whose explanation defies intuition, and a true logical inconsistency.
%
\section{The Wang experiment}\label{sec:twe} 
\begin{figure}[h]
\centering
\includegraphics[width=0.7\textwidth]{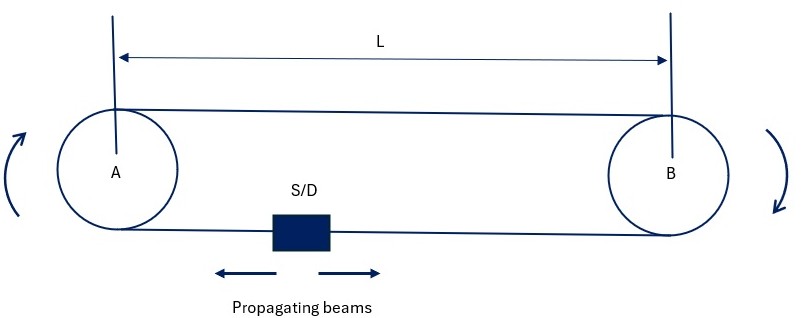}
\caption{Wang interferometer. Two beams are generated at the source (S) and then detected after returning to the detector (D). The $S/D$ device is fixed to the fiber optic medium, which is assumed to have refractive index $n=1$ and is set in motion with linear speed $\nu$ by the rotating wheels $A$ and $B$.}\label{fig1}
\end{figure}

%
Ruyong Wang et al. \cite{pWan03,pWan04} carried out an experiment that they conceived as a modified Sagnac interferometer (Fig. \ref{fig1}).
In the Sagnac interferometer, a phase shift is produced between two light beams by rotation. In contrast, in the Wang experiment, the phase shift is obtained by linear motion using a device they called \textit{fiber optic conveyor} (FOC) to distinguish it from the usual fiber optic gyroscope (FOG).

The phase shift at $D$ is owing to the time difference in the arrival times of the beams and is readily obtained in the laboratory reference frame (LAB frame) as,
\begin{eqnarray}
    \Delta t &=& \frac{4\nu L}{c^2}
\end{eqnarray}
where $t$ is the time of the LAB frame.
According to relativity, the proper time difference indicated by a clock moving with $S/D$ is
\begin{eqnarray}
    \Delta \tau &=& \frac{\Delta t}{\gamma}   
\end{eqnarray}
where the gamma factor is $\gamma=1/\sqrt{1-(\nu/c)^2}$.
Since, to first order in $v/c$, $\gamma\approx 1$, we can ignore the time dilation effect for this experiment.
%
\section{The Inconsistency Claim} 
\begin{figure}[h]
\centering
\includegraphics[width=0.8\textwidth]{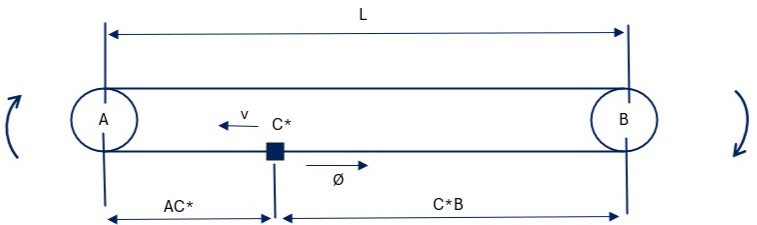}
\caption{The clock $C^*$ is stationary with respect to the fiber optic and moving with speed $\nu$ with respect to the laboratory reference frame}\label{fig2}
\end{figure}

%
According to the authors of Ref. \cite{pSpa25}, it is well-known that the Lorentz transformation fails to describe light propagation along a closed contour.
To prove their claim, they analyze the Wang interferometer and calculate the time it takes the counter-propagating photon $\phi$ to complete a round-trip, as measured by a clock $C^*$ fixed to the fiber optic and moving with speed $\nu$ with respect to the LAB frame (Fig. \ref{fig2}).

\begin{figure}[h]
\centering
\includegraphics[width=0.8\textwidth]{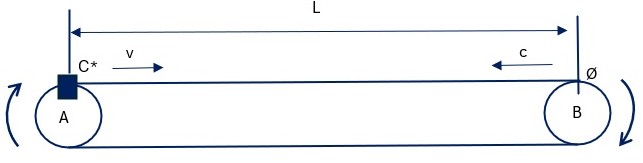}
\caption{In the LAB frame, the two distant events ``the clock is at $A$'' and ``the photon is at $B$'' are simultaneous.}\label{fig3}
\end{figure}

The time registered by $C^*$ is calculated from the LAB frame.
Assuming an initial condition where the distances of the clock $C^*$ from $A$ ($AC^*$) and $B$ ($C^*B$) are such that when the clock reaches $A$, simultaneously the photon $\phi$ reaches $B$.
Then, if $t$ is the laboratory time,
\begin{eqnarray}
    AC^* &=& \nu\,t_{out}\\
    C^*B &=& c\,t_{out}   
\end{eqnarray}
Since $AC^*+C^*B=L$, we readily obtain,
\begin{eqnarray}
t_{out}=\frac{L}{c+\nu}\label{eq:tout}
\end{eqnarray}
Disregarding the curved sections corresponding to the pulleys at $A$ and $B$, the configuration at $t=t_{out}$ is depicted in Fig. \ref{fig3}.

At $t=t_{ret}$, the clock and the photon reunite on the upper section; $t_{ret}$ must satisfy the equation $L=\nu\,t_{ret} + c\,t_{ret}$, thence,
\begin{eqnarray}
    t_{ret} &=&\frac{L}{c+\nu}\label{eq:tret}
\end{eqnarray}
The total time of the photon round-trip, according to the LAB frame time, is
\begin{eqnarray}
    T_{LAB} &=& t_{out} + t_{ret}= \frac{2L}{c+\nu}
\end{eqnarray}
The proper $T$ time registered by the clock $C^*$ is,
\begin{eqnarray}
    T &=& \frac{T_{LAB}}{\gamma}= \frac{2L}{\gamma\,(c+\nu)}=\frac{2L}{\gamma\,c\,(1+\beta)}\label{eq:T1}
\end{eqnarray}
where $\beta=\nu/c$.

The presumed inconsistency is revealed when we investigate the total distance covered by the photon in the fiber medium according to the proper time (\ref{eq:T1}) and the fact that the speed of light is $c$ irrespective of the observer.
Since the clock $C^*$ is stationary with respect to the fiber optic, we expect that after completing its journey back to $C^*$, the photon should have traveled the entire length of the fiber optic medium.
However, we find that the distance traveled by the photon is,
\begin{eqnarray}
    c\,T &=& \frac{2L}{\gamma(1+\beta)}< 2\gamma L
\end{eqnarray}
Where $2\gamma L$ is the total fiber optic proper length since the fiber moves with speed $\nu$ between the centers of the stationary pulleys and $AB=L$.

The difference
\begin{eqnarray}
   \delta d_\phi  &=& 2\gamma L - cT= 2L\gamma - \frac{2L}{\gamma(1+\beta)}=2\gamma L \beta\label{eq:mp1}
\end{eqnarray}
is the central point of the inconsistency argument.

According to Spavieri et al., $\delta d_\phi$ is a ``missing path'' which implies a ``time gap'' $\delta t=\delta d_\phi/c$ constituting an unacceptable breach in spacetime continuity as observed in the reference frame in which the clock $C^*$ remains at rest.
%
\section{Analysis of the ``missing path'' and ``time gap''} 
In Ref. \cite{pSpa25}, the authors introduce two reference frames, $S''$, at rest with respect to the lower section of the fiber, and $S'$, at rest in the fiber's upper section.
They correctly reckoned the times $T''_{out}$ and $T''_{ret}$ corresponding to time as measured in the $S''$ reference frame.

\begin{figure}[h]
\centering
\includegraphics[width=0.8\textwidth]{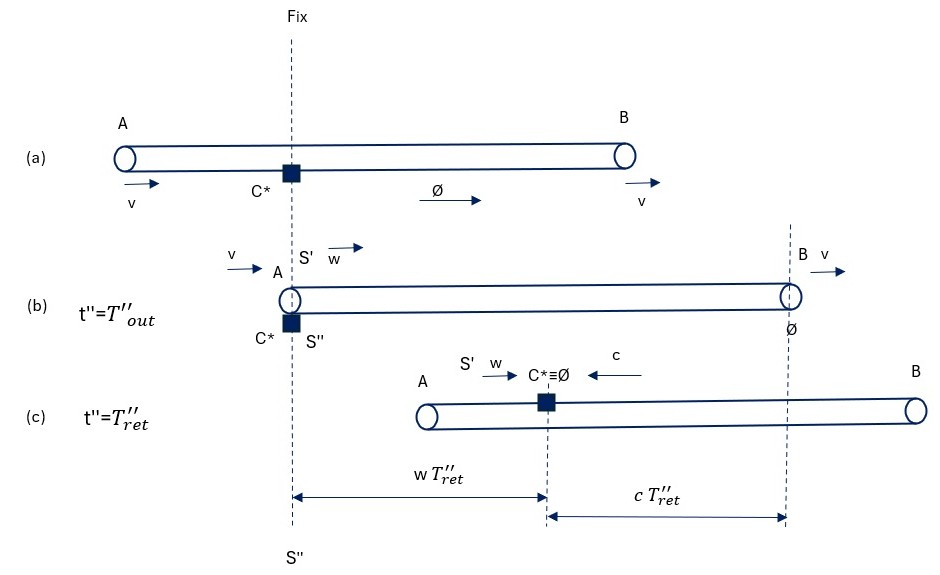}
\caption{Wang interferometer as seen from $S''$ where the lower fiber section is instantaneously at rest.}\label{fig4}
\end{figure}

They start with an initial configuration such that, as observed from $S''$, after the elapsed time $T''_{out}$, the clock reaches $A$ at the same time the photon arrives at $B$ (Figs. \ref{fig4} (a) and (b)).

According to the relativistic addition of velocities, $S'$ is moving with speed $w$ with respect to $S''$, where
\begin{eqnarray}
    w &=& \frac{2\nu}{1+\beta^2}
\end{eqnarray}
\\
Evaluating the elapsed clock's proper time $T''_{out}+T''_{ret}/\gamma_w$, we find (\ref{eq:T1}) once more, confirming the previous result by a different method,
\begin{eqnarray}
    T &=& T''_{out}+\frac{T''_{ret}}{\gamma_w}=\frac{2L}{\gamma c (1+\beta)}\label{eq:T2}
\end{eqnarray}
Where,
\begin{eqnarray}
    T''_{out} &=& \frac{L}{\gamma c}\label{eq:t2out}\\
    T''_{ret} &=& \frac{L}{\gamma c (1+\frac{w}{c})}\label{eq:t2ret}\\
    T'_{ret}  &=&\frac{T''_{re}}{\gamma_w}=\frac{L}{\gamma c}\frac{(1-\beta)}{(1+\beta)}\label{eq:t1ret}
\end{eqnarray}              
Where,
\begin{eqnarray}
    \frac{1}{1+\frac{w}{c}} &=& \frac{1+\beta^2}{(1+\beta)^2}\label{eq:1+wc}\\
    \gamma_w  &=&\frac{1}{\sqrt{1-(\frac{w}{c})^2}}=\frac{1+\beta^2}{1-\beta^2}\label{eq:gw}
\end{eqnarray}
%
\subsection{Missing path according to Spavieri et al.}
Here we reproduce the core of the argument presented in \cite{pSpa25}. Under the title: \textbf{Spatial ground distance covered by the photon according to the LT}, the authors explain that the distance covered in $S''$ according to (\ref{eq:t2out}) is,
\begin{eqnarray}
L''&=& c\,T''_{out}=\frac{L}{\gamma}
\end{eqnarray}
In the return trip on the upper fiber section, the photon is in the system $S'$, and the distance covered is, according to (\ref{eq:t1ret}),
\begin{eqnarray}
    L' &=& c\,T'_{ret}=\frac{L}{\gamma}\frac{(1-\beta)}{(1+\beta)}
\end{eqnarray}
The total distance covered by the photon in the fiber medium is,
\begin{eqnarray}
    L''+L' &=& \frac{L}{\gamma} + \frac{L(1-\beta)}{\gamma(1+\beta)}= \frac{2L}{\gamma(1+\beta)}\label{eq:L1p+Ll2p}
\end{eqnarray}
The missing path paradox arises because the total proper length of the fiber is $2\gamma L$, so we have a missing path that the photon has not covered,
\begin{eqnarray}
    \delta d_\phi &=& 2\gamma L - (L'+L'')=2\gamma L\beta\label{eq:mp2}
\end{eqnarray}
This result confirms again the value (\ref{eq:mp1}) found by another method.
\subsection{Explication of the ``missing path paradox''} 
According to Spavieri et al. \cite{pSpa25}, owing to the invariance of the speed of light, the time required by the photon to cover the whole fiber's proper length is,
\begin{eqnarray}
    (T)_E = \frac{2\gamma L}{c}
\end{eqnarray}
Since $(T)_E$ is greater than the observed time $T$ given by (\ref{eq:T1}) and (\ref{eq:T2}), the Lorentz transformation purportedly fails to provide a consistent explanation.
Spavieri et al. attribute this problem to the relativity of simultaneity.

However, we shall see that the correct application and interpretation of the Lorentz transformation (\ref{eq:LT}) proves that the total time the photon was in motion is indeed $(T)_E$ and that the missing path is inexistent.

Ironically, it is the fact that simultaneity is relative and not absolute that renders the correct explication of the missing path paradox, instead of producing it.

Let us set two reference systems $S'(t',x',y',z')$ and $S''(t'',x'',y'',z'')$ as in \cite{pSpa25}; their origins coincide when the clock $C^*$ is at $A$ (Fig. \ref{fig4} (b)).
The Lorentz transformation is given by,
\begin{equation}\label{eq:LT}
\left.
\begin{array}{ccrl}
    t' &=& \gamma_w&(t''-\frac{wx''}{c^2})\\
    x' &=& \gamma_w&(-w t''+ x'')
\end{array}
\right\}
LT
\end{equation}\\
We have two relevant events, the first one is the event $CA\equiv$``the clock is at A''.
The second one is $PB\equiv$``the photon  is at B.''
We recall here that, according to the theory of relativity, only the individual events are part of objective reality, but their eventual simultaneity is not.

The coordinates $(t'',x'')$ and $(t',x')$ of these events in $S''$  and $S'$ are
\begin{equation}\label{eq:lt0}
\begin{array}{ccrccccccclcl}
   CA &: & ( & 0 &,& 0         &)_{S''}   &\xrightarrow{LT} & ( & 0                       &,& 0                &)_{S'}\\
   PB &: & ( & 0 &,& L/\gamma  &)_{S''}   &\xrightarrow{LT} & ( & -\gamma_w wL/\gamma c^2 &,&\gamma_w L/\gamma &)_{S'}
\end{array}
\end{equation}
At $t''=t'=0$ the clock $C^*$ is abruptly forced to pass from the system $S''$ to $S.'$  
For the argument's sake, it is irrelevant whether we choose to use another clock already at rest in $S'$, as in Ref. \cite{pSpa25}, to reckon time intervals.

As shown by (\ref{eq:lt0}),  in $S''$, the events $CA$ and $PB$ have the same time coordinates $t''_{CA}=t''_{PB}=0$ (they are simultaneous), but in $S'$ we find that $t'_{CA}=0\neq t'_{PB}=-\gamma_w wL/\gamma c^2$ (they are not simultaneous).

Therefore, when the clock is forced to change reference frames at $t'=0$, we find that the event ``$PB\equiv\text{the photon is at B}$'' has already taken place in $S'$ at $t'_{PB}=-\gamma_w wL/\gamma c^2< 0$, before the clock is forced to change systems.

Then, if simultaneity is not absolute, as relativity and the Lorentz transformation require, when $C^*$ passes to $S'$, it finds that, in that reference frame, the photon departed from $B$ in the past and, at $t'=0$, has already traveled the distance
\begin{eqnarray}
    c|t'_{PB}| &=& c|-\gamma_w wL/\gamma c^2|= 2\gamma L\beta= \delta d_\phi
\end{eqnarray}
This shows that there is indeed no missing path because the total covered length is given by the distance traveled in $S''$, equal to $L''$, plus the total distance traveled in $S'$, equal to $L'$(after the clock changed frames), plus $\delta d_\phi$ (before the clock changed frames), 
\begin{eqnarray}
    L''+L'+\delta d_\phi &=& 2\gamma L
\end{eqnarray}
Accordingly, the total time the photon was in motion is given by,
\begin{eqnarray}
    |t'_{PB}| + T &=& \frac{\gamma_w w L}{\gamma c^2} + \frac{2L}{\gamma c (1+\beta)}\\
    &=& \frac{2\gamma L}{c}\\
    &=& (T)_E
\end{eqnarray}
%
\section{Conclusions} 
%
We have demonstrated that Spavieri et al.'s ``missing path paradox'' arises only if we assume that relative simultaneity is false; therefore, it cannot prove its falsity.

From the point of view of the stationary clock $C^*$, the correct explication requires the description from two distinct inertial frames, $S''$ and $S'$, and the use of the Lorentz transformation taking into account relative simultaneity, in which case no contradiction or inconsistency arises.
The need to introduce two inertial reference frames, $S''$ and $S'$, happens because the clock's frame is noninertial.
Many paradoxes arise when we naively apply our ``inertial intuition'' to noninertial reference frames, and the missing path paradox is no exception.

When the description of the entire phenomenon is made from only one inertial system, for instance, from the laboratory reference frame, or $S''$, there seems to be a lost ``time gap'' or ``missing path'', thence the paradox.\footnote{Another explanation of the paradox is given in Ref. \cite{ppMamone25}.}

The proof presented in \cite{pSpa25}, pretending to prove the inconsistency of relative simultaneity, involves circular reasoning because, by refusing to accept its implications, the alleged proof implicitly assumes what is pretended to be proved.

Intuition is a good guide when it is supported by strict logic and rigorous rational thinking, but intuition and ``common sense'' alone can be deceiving, as modern physics, founded in relativity and quantum mechanics, has taught us.
%
{\small
\section*{Author's Note} 
In his reply \cite{pSpa26} to this paper, Spavieri rejected our arguments based on:
\begin{itemize}
    \item The claim that we obtained $(T)_E$ by arbitrarily adding the extra time interval $\delta t'=t'_{PB}=\gamma_w wL/\gamma c^2$ to the observed $T$ registered by the clock.
    \item $(T)_E$ is different from the time $T$ registered by clock $C^*$, therefore it has no physical meaning.
\end{itemize}

In the first point, the fact of the matter is that $t'_{PB}$ is not an ad hoc ``deus ex machina'' artifact fabricated only to obtain a convenient $(T)_E$ value, as claimed by Spavieri, but a concrete prediction of the Lorentz transformation, as proved by (\ref{eq:lt0}) which gives the necessary value $(T)_E=T+t'_{PB}$ for the correct relativist interpretation.

Regarding the second point, the statement that the experimental outcome contradicts relativity because $(T)_E$ is not the observed time $T$ registered by the clock $C^*$ constitutes a sophism because, according to relativity, the elapsed time $(T)_E$ is not supposed to be reckoned by $C^*$, so it cannot contradict the theory.

Spavieri's argument, sustaining that $(T)_E$ must be equal to $T$, is valid only because he assumes that the events $CA$ and $PB$ are simultaneous in $S'$ because they are simultaneous in $S''$, contradicting the relativist prediction (\ref{eq:lt0}).
Thus, in his second point, rather than refuting our argument, he restates his original reasoning, confirming its circularity: relative simultaneity leads to inconsistencies only if we assume it is false.

The problems claimed by advocates of absolute simultaneity arise from missing the point that, if action at a distance is non-existent, there is no fact of the matter whether two distant events are simultaneous, i.e., distant simultaneity is not part of objective physical reality and can be observer dependent.
That is why, despite unintuitive paradoxes, neither a logical contradiction nor an empirical inadequacy has ever been found contradicting its relative character.
}
%
%
\bibliography{zRelativity}
%
\end{document}